\documentstyle[prl,aps,epsf,multicol]{revtex}
\begin{document}
\draft
 
\title{Luttinger Liquid Physics in the Superconductor Vortex Core} 
\author{Ashvin Vishwanath$^{1, 2}$ and T. Senthil$^2$}
\address{$^1$Physics Department, Princeton University, NJ 08544  \\
$^2$Institute for Theoretical Physics, University of California,
Santa Barbara, CA 93106--4030
}

\date{\today}
\maketitle



 
\begin{abstract}
We study several aspects of the structure of vortices in 
conventional $s$-wave Type $II$ superconductors. It is well-known 
that there are 
low energy quasiparticles bound to the core of a vortex. 
We show that under certain conditions, these quasiparticles form a 
degenerate Fermi gas with a finite density of states at the Fermi energy. 
In three dimensional superconductors, the result is a one dimensional 
Fermi gas of quasiparticles bound to the core of a vortex line. 
As is usual in one dimensional systems, 
interactions between the quasiparticles lead to Luttinger liquid 
behaviour. This may be probed through STM tunneling into the vortex core. 
We further suggest that a novel Peierls-type instability in the shape of
the vortex line may develop due to interactions between the quasiparticle gas and 
fluctuations in the vortex line shape.  
 
\end{abstract}
\vspace{0.15cm}

\begin{multicols}{2}

\section{Introduction}
\label{Intro}
The electronic properties of Type $II$ superconductors in the mixed
phase are known to have several interesting features. In some
pioneering work, Caroli et. al.\cite{cmdg} predicted theoretically the
existence of low energy quasiparticle states bound to the core of a
single isolated vortex in an $s$-wave superconductor.  These low lying
states are free to move parallel to the vortex, and so give rise to a
series of one dimensional bands (\emph{minibands}) as shown in
\emph{Fig 1(a)}. The system is still gapped, though the energy gap to the lowest
excitation (the minigap) is typically 
much smaller than the bulk gap. Subsequent theoretical work
\cite{bardeen,kp,gs} helped to put this prediction of low energy quasiparticle states bound to the vortex core on 
firm 
ground.
Striking experimental evidence for the existence of these bound states
of the vortex was provided by the scanning tunneling microscopy (STM)
studies of Hess et al \cite{hess}. These experiments were able to
image the vortex quasiparticle states as a function of their energy,
and agreed well with theory.

Most theoretical work on these low energy quasiparticle states in the vortex core has neglected the effect of the 
Zeeman coupling between the magnetic field and the spin of the
quasiparticles. However as discussed in Ref.\cite{brun}, and as we argue below, the Zeeman coupling plays an 
important 
role in determining the low temperature properties
of the system. The Zeeman coupling splits the quasiparticle energy levels, and the minigap is decreased since the 
energy 
of one spin species is lowered towards the Fermi energy. For sufficiently strong Zeeman splitting,
the minibands of one spin species will be brought down below the
Fermi energy, and a filled Fermi sea of spin polarised
quasiparticles is formed. We show that the magnetic fields needed in
typical materials to form this degenerate quasiparticle system are not
large, and could be much smaller than $H_{c2}$.

We consider the effect of quasiparticle-quasiparticle
interactions (ignored in the BCS mean field theory) on the low energy
properties of the quasiparticles in the core of the vortex.  The
quasiparticles are bound in the direction perpendicular to the vortex
line but are free to move along it, thus providing an interesting
realization of a one dimensional system inside the superconductor.  It
is well-known in the theory of normal metals that interaction effects
are dramatic in one dimension: the generic ground state of the
interacting electron system is not a Fermi liquid but a different
beast, the Luttinger liquid. We therefore focus attention primarily in
the regime of magnetic fields well below $H_{c2}$ where the vortices
may be treated in isolation.  Are some of the striking properties of
interacting 1D Fermi systems, such as Luttinger liquid physics, also
present in the vortex quasiparticle system? The answer is
no, if the interactions and the Zeeman coupling are weak.
The presence of the (mini)gap implies that the T=0 state of this
system is quite insensitive to weak interactions. However if the
Zeeman energy is large enough to start filling the miniband, the
system is gapless - interaction effects are then crucial, and we argue
that the system is correctly described as a spin polarised Luttinger liquid. We then find that the 
interaction strength \emph{g} which controls the Luttinger liquid
exponents is a function of the miniband filling, and in
principle could even be negative.  Luttinger liquids have several
interesting properties \cite{luttinger} - the fermion correlation functions are power
law with anomalous exponents and the single particle density of states at the Fermi
points vanishes as a power law. There have been a small number of
experimental realizations of Luttinger liquid behaviour in systems
such as 1D semiconductor wires \cite{tarucha}, fractional quantum hall effect edge states \cite{fqhe} and carbon nanotubes \cite{balents}. We show that a
Luttinger liquid can be realised under appropriate conditions in the
vortex core.  Tunneling into a Luttinger liquid leads to a 
characteristic power law tunneling conductance, and we propose an experiment
involving Scanning Tunneling Microscopy (STM) measurements on the vortex core as a probe of Luttinger liquid 
behaviour 
of the vortex quasiparticles.

The quasiparticles also interact with the collective modes of the
vortex, which can have interesting consequences. In particular, if
inter-vortex interactions (which become increasingly important as we
increase the field towards $H_{c2}$) are sufficiently strong, we find that a vortex analog of the Peierls effect 
could 
occur. The vortex line spontaneously
undergoes a periodic modulation of its profile to lower the energy by opening up a gap in the quasiparticle 
spectrum.
	 
The rest of this paper is organised as follows. In Section
\ref{zeeman}, we discuss in a little more detail the core states of an
isolated vortex, and the formation of the degenerate quasiparticle gas
in the presence of the Zeeman splitting. In Section \ref{LL} we show
that interactions between the quasiparticles can lead to them forming
a Luttinger liquid, with a varying Luttinger liquid exponent and
consider how this state may be observed in STM experiments. In Section
\ref{VPE}, we couple the quasiparticles to the vortex collective modes
and discuss in particular the possibility of a vortex Peierls transition when inter-vortex interactions are 
present. 
Section \ref{disc}
discusses the validity of some of our approximations and highlights
directions where more work needs to be done.

\section{Vortex Quasiparticle States in the Non-Interacting Limit}
\label{zeeman}
In this section we review the spectrum of the vortex core states and discuss the effect of the Zeeman 
coupling on it. We begin with the BCS model Hamiltonian in the presence of an arbitrary magnetic field 
$\vec{B}=\vec{\nabla} \times \vec{A}$ and pair potential $\Delta(r)$ (we will include the Zeeman term 
later):

\begin{eqnarray}
H_{BCS} & = & \int d^dx ~ c^{\dagger}_{\sigma}(x) \varepsilon [-i\nabla -eA(x)]  c_{\sigma}(x) + \nonumber \\
&  &  c^\dagger_{\uparrow} (x)\Delta(x)c^\dagger_{\downarrow} (x)  +  c_{\downarrow} (x)\Delta^*(x) 
c_{\uparrow} (x)
\label{HBCS} 
\end{eqnarray}
where $\varepsilon[p]$ is the kinetic energy measured from the chemical potential. For simplicity, we will assume a 
quadratic energy dispersion $\epsilon(p) = \frac{p^2}{2m}-E_F$. It is convenient to rewrite the Hamiltonian using 
the 
change of variables

\begin{eqnarray*}
d_{1}(x) & = & c_{\uparrow}(x)\\
d^{\dagger}_{2}(x) & = & c_{\downarrow}(x)
\end{eqnarray*}

When written in these variables, there are no anomalous terms in the Hamiltonian: 

\begin{eqnarray}
H_{BCS} & = &
\int d^dx  \left( \begin{array}{cc}
	d_{1}^{\dagger}(x) & d_{2}^{\dagger}(x) \end{array} \right)
H
\left( \begin{array}{c}
	d_{1}(x)\\
	d_{2}(x) \end{array} \right)\\
H & = & \left( \begin{array}{cc}
      \varepsilon[-i\nabla-eA(x)] & \Delta^{*}(x)\\
      \Delta (x) & -\varepsilon[i\nabla-eA(x)] \end{array} \right)
\end{eqnarray}

Physically, the number of $d$-particles corresponds to the $z$-component of the electron spin.  
The absence of anomalous terms in the $d$ representation thus reflects conservation of the 
$z$-component of the electron spin in the BCS Hamiltonian of Eqn. \ref{HBCS}.

The Hamiltonian is diagonalised by going over to new quasiparticle operators that are defined as:

\begin{eqnarray*}
\gamma_{1\alpha}^\dagger & = & \int(u_{\alpha}(x)d_{1}^\dagger(x)+v_{\alpha}(x)d_2^\dagger (x))d^dx 
\\
\gamma_{2\alpha}^\dagger & = & \int(v_{\alpha}^*(x)d_{1}^\dagger(x)-u^*_{\alpha}(x)d_2^\dagger 
(x))d^dx
\end{eqnarray*}
The functions $u_\alpha(x)$ and $v_\alpha(x)$ are found by solving the following eigenvalue equation 
(with $E_\alpha \ge 0$): 
\begin{equation}
H
\left( \begin{array}{c}
	u_{\alpha}(x) \\
	v_{\alpha}(x) \end{array} \right)
= E_{\alpha}
\left( \begin{array}{c}
	u_{\alpha}(x)\\
	v_{\alpha}(x) \end{array} \right)
\label{bdg}
\end{equation}
	
	This is just the Bogoliubov deGennes (BdG) equation for the quasiparticle states. The BdG Hamiltonian can 
be 
recast in a compact matrix notation as:
$$
H = \varepsilon[-i\nabla {\bf {\tau^z}}-eA(x)]{\bf \tau^z} + \Delta^*(x) {\bf{\tau^+}} + \Delta (x) {\bf \tau^-}
$$
where $\vec{\tau}$ are the $2 \times 2$ Pauli matrices. In terms of the quasiparticle operators $\gamma$, the 
Hamiltonian is 
simply:
\begin{equation}
H_{BCS}=
\sum_{\alpha} \left( \begin{array}{cc}
	\gamma_{1\alpha}^{\dagger}(x) & \gamma_{2\alpha}^{\dagger}(x) \end{array} \right)
\left( \begin{array}{cc}
      E_{\alpha} & 0\\
      0 & -E_{\alpha} \end{array} \right)
\left( \begin{array}{c}
	\gamma_{1\alpha}(x)\\
	\gamma_{2\alpha}(x) \end{array} \right)
\end{equation}

The ground state has all of the negative energy states filled so 
$\gamma^\dagger_{2\alpha}|0\rangle=0$; all positive energy states are unoccupied 
$\gamma_{1\alpha}|0\rangle=0$.  Note that a spin $\uparrow$ excitation is created by ($\gamma^\dagger_1$) while a 
spin 
$\downarrow$ excitation is created by ($\gamma_2$) acting on the ground state. 

After these generalities, let us specialise to the quasiparticle states bound to the vortex line in a 
superconductor. Consider an isolated, straight vortex oriented along the z-axis, sitting in a clean, conventional 
Type II 
superconductor. This is realised if the applied magnetic field is just above $H_{c1}$. The vortex bound state 
energy 
levels are obtained by solving Eqn. \ref{bdg} with the appropriate gap profile ($\Delta(r)=|\Delta(r)|e^{i \phi}$, 
where 
$\phi$ is the azimuthal angle about the vortex).  It is useful to perform the singular gauge transformation that 
makes 
$\Delta$ real everywhere (the London gauge):

\begin{eqnarray}
 \left( \begin{array}{c}
      u'\\
      v' \end{array} \right) & = &
\left( \begin{array}{c}
	e^{+i\frac{\phi}{2}} u\\
	e^{-i\frac{\phi}{2}} v \end{array} \right) \\
H' \left( \begin{array}{c}
       u' \\
       v' \end{array} \right) & = &
E \left( \begin{array}{c}
      u'\\
      v' \end{array} \right) \\
H' = 
	[\frac{1}{2m}(-i\nabla {\bf {\tau^z}}-eA -\frac{\hat{\phi}}{2r})^2 & - & E_F]{\bf {\tau^z}} + 
|\Delta(r)|{\bf {\tau^x}}
\end{eqnarray}
The transformed quasiparticle wavefunctions $ \left( \begin{array}{c} u'\\ v' \end{array} \right)$ obey 
antiperiodic 
boundary conditions to keep $ \left( \begin{array}{c} u\\ v \end{array} \right)$ single valued.

\end{multicols}
\begin{figure}
\epsfxsize=7in
\centerline{\epsffile{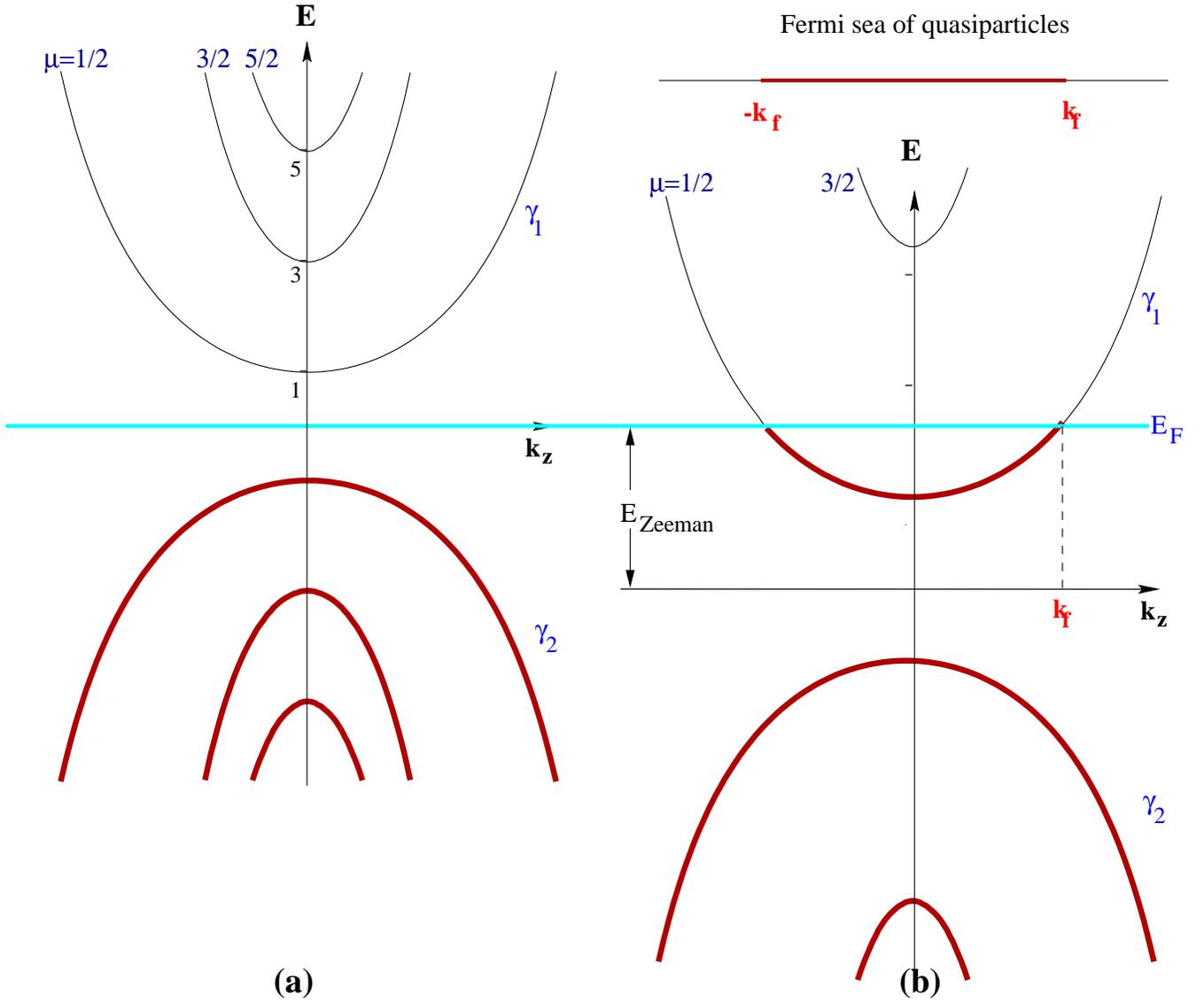}}
\vspace{0.15in}
\caption{Energy level structure of the states bound to the core of a vortex (a) Ignoring the effect of the magnetic field. The negative energy states are occupied and shown with the thick line. Energy is measured in units of the minigap energy ($\approx \frac{\Delta^2}{E_F}$)  and 
(b) In the presence of a sufficiently strong magnetic field. The Zeeman splitting causes the $\mu = \frac12$ miniband to start being occupied and a degenerate Fermi sea is formed.}
\vspace{0.15in}
\label{vxcr}
\end{figure} 

\begin{multicols}{2}
We 
first review the results of Ref. \cite{cmdg} on the structure of an isolated vortex. In Ref. \cite{cmdg} the effect 
of the 
magnetic field on the low energy quasiparticles is not included. This is only a good approximation for an isolated 
vortex 
in an extreme Type II superconductor; then the flux enclosed in the spatial region where the low energy states are 
bound is a 
small fraction of the flux quantum, and hence the magnetic field has little effect on these states. Subsequently we 
will 
see how these results need to be modified on including the effect of the magnetic field. The low energy spectrum of 
the vortex core states then is :
\begin{equation} 
E^0(\mu,k_{z})=\mu \left(\frac{c\Delta_\infty}{k_{F} \xi}\right) 
\left[1-\frac{k_{z}^{2}}{k_{F}^{2}}\right]^{-\frac12}
\label{CMdG}
\end{equation}
and is shown in Fig 1(a). The corresponding quasiparticle wavefunctions in cylindrical coordinates take the form 
\begin{eqnarray}
 \left( \begin{array}{c}
      u'_{\mu k}(r,\phi,z)\\
      v'_{\mu k}(r,\phi,z) \end{array} \right) & = &
e^{i\mu \phi}e^{ik_zz}\left( \begin{array}{c}
	f_{\mu k_z}(r)\\
	g_{\mu k_z}(r) \end{array} \right)
\label{states}
\end{eqnarray} 
where $\mu \in \{\frac12, \frac32, \frac 52 ...\}$ appears as an angular momentum in the wavefunctions and is a half 
integer due to the anti-periodic boundary conditions, $k_z$ is the momentum along the vortex line, $c$ is a 
constant of 
order one and $\xi$ is the zero temperature coherence length. We have denoted by $\Delta_{\infty}$ the value of the 
gap 
very far away from the vortex. The $k_z$ dependence of the energy comes from the motion along the vortex line. The 
energy scale of the minigap is $\frac{\Delta_{\infty}^2}{E_F}$ and is much smaller than the bulk gap 
$\Delta_{\infty}$. 
The radial wave funcions $f$ and $g$ are mainly confined to within a radius $\xi$ (the coherence length) about the 
vortex for the low energy states ($E(\mu ,k_z)\ll \Delta_\infty$).

Now consider the effect of the magnetic field $\vec{B}=B\hat z$. We assume though that the field is much less 
than $H_{c2}$ so that the vortices are well separated. Two effect occur \cite{brun}: (a) the Zeeman term splits the 
energy levels and (b) there is an orbital effect. We consider these in turn, starting with the orbital effect.

For the magnetic fields of interest, the inter-vortex separation is typically smaller than the penetration depth so 
that 
the field in the superconductor is fairly uniform and equal to the external field ($=B\hat{z}$, and we take 
$\vec{A}(\vec{x})=\frac{B}{2}(-y,x,0)$). 
In the presence of the vector potential additional terms are generated in the BdG Hamiltonian which in the London 
gauge 
take the form:
\begin{eqnarray}
\delta H & = & \delta H_1 + \delta H_2 \\
\delta H_1 & = & \frac{|e|}{2m}(A(x)\cdot p + p\cdot A(x))\\
\delta H_2 & = & \frac{(eA(x))^2}{2m}{\bf \tau^z} + \frac{|e|B}{2m}{\bf \tau^z}
\end{eqnarray}
$\delta H_2$ may be neglected since the first term is small compared to the minigap for fields well below $H_{c2}$ 
and 
the second term only makes a small shift of the Fermi energy. 
So we are left with $\delta H_1$ which can be written as:  
\begin{equation}
\delta H_1 = \frac{|e|}{2m}B \vec{r} \times \vec{p} = \frac{|e|}{2m}B L_z
\end{equation}
This does not affect the eigenstates (\ref{states}) but shifts their energy by

\begin{equation} 
E(\mu,k_z,B) = E^0(\mu,k_z) + (\frac{|e|B}{2m})\mu
\end{equation}
The orbital effect therefore increases the magnitude of energy of the quasiparticle states.

The Zeeman term takes the form  
\begin{eqnarray}
H_Z & = & \frac{-g\mu_B B}{2}\int 
(c^{\dagger}_{\uparrow}(x)c_{\uparrow}(x) - c^{\dagger}_{\downarrow}(x)c_{\downarrow}(x)) d^dx \\
	& = & \frac{-g\mu_B 
B}{2}\int \left(d_{1}^{\dagger}(x)d_{1}(x)+d_{2}^{\dagger}(x)d_{2}(x)\right) d^dx \\
	& = &  \frac{-g\mu_B B}{2}\sum_\alpha 
(\gamma_{1\alpha}^\dag\gamma_{1\alpha}+\gamma_{2\alpha}^\dag\gamma_{2\alpha}) 
\end{eqnarray}
(In the second line, we have dropped an irrelevant constant and $\mu_B = \frac{|e|\hbar}{2m}$ is the Bohr magneton). 
The 
Zeeman term thus behaves as a `chemical potential' for the d-particles \cite{spin}. Increasing the magnetic field 
raises 
this chemical potential. Beyond a certain field it enters the first miniband and a degenerate Fermi sea of 
quasiparticles forms as shown in Fig 1(b). The condition for this is simply:

\begin{equation}
E(\mu=\frac12,k_z=0,B) \le  \frac{g\mu_B B}{2}
\end{equation}
or equivalently
\begin{equation}
E^0(\mu=\frac12,k_z=0)\le \frac{(g-1)}{2}\mu_B B
\end{equation}

Thus a degenerate gas of spin polarised fermions (in this case all spin $\uparrow$) can form due to the Zeeman 
coupling 
if the magnetic field exceeds $H_{cZ}=\frac2{\mu_B}E^0(\mu=\frac12,k_z=0) $ (assuming $g=2$).  Note that it is {\em only} 
the first 
miniband that can be brought below the chemical potential. For $\mu>\frac12$, the orbital effect wins over the 
Zeeman 
splitting, and the levels do not cross the chemical potential.

The magnetic fields needed to begin filling the miniband states are very reasonable, since the minigap energy is 
typically small. To illustrate this point we consider the case of NbSe$_2$ for which numerical calulations of the 
vortex 
quasiparticle spectrum are available \cite{gs}. NbSe$_2$ has a superconducting transition temperature of $T_{c}=7.2 
K$, 
and numerical calculations find the minigap to be $0.4$ Kelvin. The magnetic field required to close the minigap is 
[$H_{cZ}$] is $1.3$ Tesla (we have assumed $g=2$) while $H_{c2}$ is larger at $3.2$T for this 
material. Better candidate materials will have smaller ratios of $\frac{H_{cZ}}{H_{c2}}$ than NbSe$_2$.

In passing we note that the Zeeman splitting gives rise to quasiparticles even at zero temperature, and in doing so 
`melts' some of the 
condensate at the centre of the vortex. The new profile $\Delta(r)$, can be calculated from the self consistency 
equation. This is analogous to the Kramer-Pesch effect \cite{kp}, where temperature does the job of melting the 
condensate and expanding the vortex. We shall not take into account the effects of the changing gap profile on the 
quasiparticle states, as they are expected to be small.

\section{Interaction Effects and Luttinger liquid Formation}
\label{LL}

Interactions between fermions in clean 1D systems have a dramatic effect resulting in Luttinger liquid behaviour. 
Here 
we shall consider the effect of interactions between the quasiparticles in the vortex core that form the 1D Fermi 
gas. A few comments on 
the nature of these interactions\cite{smitha} is in order. The underlying microscopic electronic system is 
characterized, at low energies,
by three qualitatively different kinds of interactions: (a) the spin singlet density-density
interaction in the particle-hole channel (b) the triplet spin density-spin density interaction in the 
particle-hole channel and (c) the interaction in the particle-particle channel. 
Conventional BCS superconductors arise when the interactions in the particle-particle channel are attractive. 
Indeed the electron gas is unstable toward the formation of Cooper pairs in the  
presence of such an attractive interaction. The BCS theory simply treats this attractive interaction 
in the particle-particle channel in a mean field approximation. One still however has to deal 
with the interactions in the particle-hole channel. (These are simply ignored in the so-called reduced BCS
Hamiltonian). Inclusion of these leads to interactions between the quasiparticles in the 
superconductor. Note that the long-ranged Coulomb repulsion between the electrons in the singlet channel 
is screened out by the condensate in the superconducting phase. Thus the residual interactions between the 
quasiparticles are short-ranged. A further source of interactions between the quasiparticles comes from the 
inclusion of fluctuations of the mean field order parameter (again usually ignored in BCS theory).

Our general conclusions are insensitive to the precise form of this short-ranged interaction
between the quasiparticles. We will therefore, for illustrative purposes, consider a particular model interaction. 
We 
add to the BCS Hamiltonian Eqn.\ref{HBCS} the ``interaction'' term
\begin{equation}
\label{interaction} 
H_{int} = \int \sum_{\sigma \sigma '}V(|x-x'|)c^\dagger_\sigma (x) c^\dagger_{\sigma'}(x') 
c_{\sigma'} (x') c_{\sigma} (x) d^3x d^3x'
\end{equation}
where $V(|x-x'|)$ is assumed to be short-ranged. To study the effect on the vortex core states, it is 
convenient to re-express this in terms of the $\gamma$ operators introduced in Section \ref{zeeman}

\begin{eqnarray}
\label{qpint}	 
 H_{int} & = & 
\sum_{\alpha\beta\alpha'\beta'}\int[-V(|x-x'|)]:(u^*_{\alpha}(x)\gamma^{\dagger}_{1\alpha}+v_{\alpha}(r)\
\gamma^{\dagger}_{2\alpha})\\
 & & \times (u_{\beta}(x)\gamma_{1\beta}+v_{\beta}^{*}(x)\gamma_{2\beta}) \nonumber \\  \nonumber
&	& \times 
(v^*_{\alpha'}(x')\gamma^{\dagger}_{1\alpha'}-u_{\alpha'}(x')\gamma^{\dagger}_{2\alpha'})\\	\nonumber
& & \times (v_{\beta'}(x') 
\gamma_{1\beta'}-u_{\beta'}^{*}(x')\gamma_{2\beta'}):d^dx d^dx'  \\ \nonumber
	&	& +\ldots +\ldots + \ldots	\nonumber
\end{eqnarray}
where the normal ordering is with respect to the superconductor vacuum, and we have written out only the first of 
the 
four terms present from different spin combinations. 

Let us study the effect of interactions when we may treat the vortices as isolated and the lowest miniband is 
occupied 
with the two Fermi points at $\pm k_f$ (so $E_{\frac{3}{2},0}>E_{Zeeman}>E_{\frac{1}{2},0}$). Notice that the Fermi 
Sea 
consists of spin up quasiparticles only. The excitation spectrum for this case contains a low energy part of 
`particle-hole' excitations in the vicinity of the Fermi points, 
and a high energy sector that involves either $\gamma_1$ quasiparticles hopping into higher minigap states, or the 
destruction of $\gamma_2$ quasiparticles from deep below the chemical potential. The latter two are gapped, with 
the gap 
being of order the minigap energy ($\approx \frac{\Delta}{k_F\xi}$). 
Since we are interested in the effects of weak interactions on the $T=0$ state of the system, we confine ourselves 
to 
the low energy sector of the problem. Formally, we can achieve this by defining a projection operator $P_\epsilon$ 
that 
only retains states close to the Fermi points, i.e. $\{\gamma_1$ quasiparticles, $\mu=\frac12$, $k_z\in 
(k_f-\epsilon,k_f+\epsilon)$ or $ k_z\in (-k_f-\epsilon,-k_f+\epsilon); \epsilon \ll k_f\}$. Now, we can project 
the interaction term 
Eqn(\ref{qpint}) down into this subspace, and it is easy to see that the only non-trivial term that remains is:

\begin{equation}
H^{P}_{int} = U\sum_{|q|<\epsilon}\rho_L(q)\rho_R(-q)
\end{equation} 

where the left (and similarly the right) density operators $\rho_L$ and  $\rho_R$ are constructed from the fermion 
operators in the usual way, $\rho_L$ for example is given by: 

$$
\rho_{L} = \sum_{|k+k_f|<\epsilon} \Gamma^\dagger_{k+q}\Gamma_k
$$

and $\Gamma_k=\gamma_{1,\frac12,k}$. 
In general, doing a quantitative calculation of $U$ for a real material is not an easy task because a 
complete knowledge of the interaction potential and quasiparticle wavefunctions is needed. We simply 
note that $U$ will be a function of the filling of the miniband ($U=U(k_f)$), and in principle can even 
be negative (attractive interaction between the quasiparticles) \cite{toy}.

For the particular interaction of Eqn. \ref{interaction} we can express $U$ in terms of  
the interaction potential and the wavefunction of the states near the Fermi points:  

\begin{equation}
U = \int_{x,y,x',y'} V(x-x',y-y',z)Q_{k_f}(x,y)Q_{k_f}(x',y')[1-e^{2ik_fz}]
\end{equation}

Here the quasiparticle `charge' $Q_{k_f}$ is defined by

\begin{equation}
Q_{k_f}(x,y)=|u_{\frac12, k_f}(x,y)|^2 - |v_{\frac12, k_f}(x,y)|^2
\end{equation}

We thus have a system of one-dimensional fermions with two Fermi points interacting through $H^P_{int}$. The 
fermions are spin-polarized (and hence, effectively, spinless).
It is well-known that such a system is correctly described not as a Fermi liquid, but rather as a
Luttinger liquid.

 Luttinger liquids are characterised by power law correlations. The exponents are controlled by the dimensionless quantity $g=\frac{U}{2\pi v_f}$ where $v_f$ is the Fermi velocity of the one dimensional system. For example, $\langle 
\Gamma^\dag (z)\Gamma(z')\rangle \approx(1/|z-z'|^\nu)$ where $z$ is the position along the vortex line and the exponent 
$\nu=(1/\sqrt{1-g^2})$. Relatedly, tunneling into the bulk of a Luttinger liquid is characterised by 
nonlinear 
I-V 
characteristics, in fact at low voltages the tunneling current-voltage relation is of the form $I\propto 
V^\nu$. This suggests that STM could be used to experimentally verify Luttinger liquid formation in the 
vortex cores under the conditions described above. At sufficiently low temperatures when the Luttinger 
liquid behaviour is dominant, we expect the tunneling conductance close to the centre of the vortex to be given by 
:
\begin{equation}
\sigma(V,x)\propto |u_{\frac12,k_f}(x)|^2 |V|^{\nu_e-1}
\end{equation}
and $\nu_e$ is the exponent for tunneling into the edge of a Luttinger liquid $\nu_e=(1+g)^\frac12 
(1-g)^{-\frac12}$ \cite{fisher}. The temperatures at which such Luttinger liquid behaviour will be 
observed is necessarily small compared to the degeneracy temperature, which for the half filled miniband of 
$NbSe_2$ is 
$\sim 0.4 $Kelvin \cite{temp}. These effects may be observed at higher temperatures by using a superconductor with 
a 
larger minigap, but this would also require larger magnetic fields to close the minigap via the Zeeman splitting.

\section{Coupling to Vortex Collective Modes and Vortex Peierls Effect}
\label{VPE}
So far we have only considered straight and rigid vortex lines, however in reality the vortex is a soft object and 
can 
undergo shape fluctuations. These collective modes of the vortex interact with the quasiparticles bound to the 
vortex 
line. As we show below the interacting quasiparticle - collective mode system is similar to the problem of 
interacting electrons 
and phonons in one dimension, which is known to undergo a Peierls transition (for commensurate filling) in which 
the 
lattice spontaneously distorts and a gap opens at the Fermi points. The gain in electronic energy by opening of the 
gap 
offsets the elastic energy cost of the lattice distortion. The electron scattering from the vicinity of one Fermi 
point 
to the other (2$k_F$ scattering) involves vanishing energy denominators and is responsible for this instability. 

We can ask if an analogous effect occurs in the degenerate 1D quasiparticle system in the vortex core interacting 
with 
the shape fluctuations of the vortex - which can give rise to 2$k_F$ scatterings \cite{lattice} \cite{umklapp}.  
Note 
that for an isolated vortex line, which is a one dimensional object it is not possible to spontaneously break the 
continuous symmetry of translations along the vortex line. 
 However in the vortex lattice if inter vortex interactions are sufficiently strong, a vortex analog of the Peierls 
effect could occur at finite temperature. This issue was raised earlier by Bouchaud  \cite{bouchead} who modelled 
the 
vortex core simply as a wire of normal electrons. However we believe this is not an adequate model to discuss this 
phenomena. A degenerate Fermi system is formed in the vortex only in the presence of the Zeeman splitting.

The collective modes of the vortex may be classified according to their angular momentum. Scalar modes scatter 
quasiparticles within the same $\mu$ miniband, while the lowest order coupling of quasiparticles to collective 
modes of 
higher angular momenta (say $l$)  will scatter quasiparticles from one miniband ($\mu$) to a different one ($\mu 
\pm 
l$) by conservation of angular momentum. Since the low energy excitations all lie in a single $\mu=\frac12$ miniband we 
only 
consider coupling of quasiparticles to the scalar collective modes of the vortex.

Scalar vortex modes modulate the  radial profile of the vortex as one moves along the vortex line. For instance, a 
periodic modulation of the vortex radius (defined say as 
when the order parameter reaches half its asymptotic value) is an example of a scalar deformation of the vortex. If 
the scalar normal modes of the vortex are labelled by $n$, let $\sigma^\dag_n(q_z)$ create the n$^{th}$ radial mode 
with momentum $q_z$ along the 
vortex line. The coupling of the vortex modes to the quasiparticles (of the partially filled miniband) can be derived from the BdG equation, or simply from symmetry is found to be:
\begin{equation}
\label{hintscalar} 
H_{int}^{scalar}=\sum_{q,k,n}g_{int}(k,q,n)\sigma_n(q)\Gamma^\dag_{k+q}\Gamma_{k} + h.c.
\end{equation}
The Hamiltonian for the scalar modes is:

\begin{equation}
H_{modes} = \sum_{n,q} E(n,q) \sigma^\dag_n(q) \sigma_n(q)
\end{equation}

Thus, scalar modes of the vortex of momentum 2$k_f$ will scatter quasiparticles from one Fermi point to the other. 
As we 
have noted previously, for an isolated vortex this does not lead to a gapped phase and the Luttinger liquid 
behaviour 
will persist. However if inter vortex interactions are sufficiently strong, then a finite temperature transition to 
a 
vortex Peierls phase can occur. Which of the modes $n_0$ has the highest transition temperature is a function of 
$g_{int}$, energy and inter vortex coupling for that mode. In the vortex Peierls phase, the quasiparticle spectrum 
is 
gapped, and we have a collective mode condensate i.e. $\langle \sigma_{n_0}(2k_f)\rangle\ne 0$ which implies a 
periodic 
modulation of the vortex profile.

  We do not attempt to predict the precise conditions for 
the 
experimental realization of this phase. Increasing the vortex density, and hence increasing intervortex 
interactions can 
raise the transition temperature, as long as the core quasiparticles do not hop between vortices and destroy the 
one-dimensional nature of the system. Within a mean field approximation, the Fermi gas always undergoes the Peierls transition. The instability is 
enhanced 
for repulsive Luttinger liquids \cite{voit}. Thermal transport measurements along the vortex line are expected to be 
sensitive 
to the vortex Peierls transition.  

The Luttinger liquid behaviour of the vortex quasiparticles would also be destroyed if pairing of quasiparticles were to occur. This would induce an additional (eg. p-wave) pairing amplitude in the vortex core, and could also lead to breaking of the rotational symmetry of the vortex line. This interesting situation was considered by Makhlin and Volovik in \cite{volovik}. Once again, the transition temperature for this instability is nonzero only in the presence of suitable inter-vortex interactions.

\section{Discussion}
\label{disc}
We have shown that a one-dimensional degenerate gas of quasiparticles
can form in the vortex core, as a result of the Zeeman coupling and
can serve as a laboratory for one dimensional physics. We showed that quasiparticle interactions drive the 
formation
of a Luttinger liquid inside the vortex core.  The interaction
strength $U$ of the Luttinger liquid was shown to be a function of the
filling, and could even be negative. Thus we have a 1D interacting Fermi system, where novel features arise due to 
the fact that the fermions involved are
superconducting quasiparticles and the `wire' confining them is a
vortex.

We now examine in more detail some of the approximations that have
been made, and the prospects for verifying the results we predict in
experiments.

One of the main approximations that we have made is to treat each
vortex as effectively isolated, which gives rise to the one
dimensional nature of the vortex quasiparticle system. As the magnetic
field is increased towards $H_{c2}$, the vortices get closer to each
other and the wavefuctions of the vortex core states start to overlap. This leads to quasiparticle hopping between 
vortices that will eventually
destroy the one dimensional nature of the system. To keep the hopping
small, we need that the separation between vortices at the magnetic
fields of interest is much larger than the coherence length. We therefore require that the field $H_{cZ}$  at  
which 
levels start to fill satisfies $H_{cZ} \ll H_{c2}$. However, the temperature scale associated with the field 
$H_{cZ}$ should 
not be too small since the physics of interest occurs only below that temperature, suggesting materials with a 
relatively high $T_c$. A possible family of candidates that meet these criteria are the borocarbides \cite{cava}. 
Another possibility is to work with a material that has
$H_{cZ}<H_{c1}$. Then even at magnetic fields just above $H_{c1}$,
when the vortices are very well isolated, a degenerate quasiparticle
liquid will be formed in the vortex cores. Elemental Niobium, that has a relatively large $H_{c1}=0.14$T is 
possibly such a 
system.

Throughout this work we have assumed that the superconductor is perfectly
clean, but in any real system disorder is always present. The
stability of the Luttinger liquid to weak disorder depends on the
value of the interaction strength $g$ \cite{giam}.  It is well-known
that there exists a critical $g_c < 0$ such that for $g > g_c$, any
disorder kills the Luttinger liquid behaviour leading instead to a
phase with localized quasiparticle excitations.  Still, for
sufficiently clean systems there is a range of temperatures for which
the properties of the system are controlled by the LL fixed point,
even though the ultimate zero temperature state may be 
localised. In that case, the scanning tunneling conductance that we
predict for the LL should be obtained in this crossover regime.

On the other hand, if $g < g_c$, the Luttinger liquid is, in fact,
stable to weak disorder\cite{giam}. Since $g$ for the vortex core
Luttinger liquid could be negative, the following interesting effect
can occur.  By varying the external magnetic field, $g$ can be varied
and the system can be tuned through a one dimensional delocalization
transition from the phase with localized quasiparticles (at small
negative $g$) to the Luttinger liquid (at large negative $g$).

The formation of the degenerate quasiparticle gas in the vortex core
can be probed by measurements of the low temperature specific heat - a
linear temperature dependence with a field dependent coefficient
should obtain at the lowest temperatures.  Once signatures of the
formation of this quasiparticle gas are observed and some of its
physical parameters (eg. the minigap and dispersion with $k_z$)
measured, it would be of great interest to look for possible Luttinger
liquid behaviour in, for instance, STM tunneling into the vortex core.

We also considered coupling the vortex quasiparticles to the
collective modes of the vortex. We find that a vortex analog of the
Peierls transition with a non-zero transition temperature could arise if
the inter-vortex interactions are strong enough.

\section{Acknowledgements}

We would like to thank K. Damle, C. Dasgupta, D. Huse, N.P. Ong, A. Melikidze, S. Sondhi and especially 
D. Haldane for useful discussions. One of us (A.V.) acknowledges support from grant NSF DMR-9809483.
T.S was supported by NSF Grants DMR-97-04005,
DMR95-28578
and PHY94-07194.

\end{multicols}

\begin{thebibliography}{x}
\bibitem{spin} Equivalently, we can define spin up and spin down quasiparticle states with positive 
energies and these are then Zeeman split in the usual manner.
\bibitem{toy} Calculations with toy models of the interaction show that the dimensionless ratio $g/v_f$ 
that controls the non-trivial Luttinger liquid exponents can be quite large i.e. O(1) which means the 
L.L. exponents would differ considerably from the free fermion values.
\bibitem{temp} For comparison, the experiments of \cite{hess} were done at $0.3$ Kelvin, but at low 
magnetic fields.
\bibitem{lattice} A distortion of 
the crystal lattice will not be energetically favourable for this system due to the 
low density of quasiparticles forming the 1D gas in the vortex core
\bibitem{umklapp} At some special wavevectors, when $4k_f=Q_z$(a reciprocal lattice vector), it is possible that 
Umklapp 
scattering drives a density wave formation, but this requires fine tuning of the filling (Zeeman splitting).
\end{thebibliography}
\end{document}